\documentclass[fleqn,usenatbib]{mnras}

\usepackage{newtxtext,newtxmath}

\usepackage[T1]{fontenc}

\DeclareRobustCommand{\VAN}[3]{#2}
\let\VANthebibliography\thebibliography
\def\thebibliography{\DeclareRobustCommand{\VAN}[3]{##3}\VANthebibliography}

\usepackage{graphicx}	
\usepackage{amsmath}	

\usepackage{color}

\usepackage{ulem} 
\usepackage{url}
\usepackage{comment}
\usepackage{threeparttable} 
\usepackage[figuresright]{rotating} 
\usepackage{booktabs}
\usepackage{array}
\usepackage{makecell}
\usepackage{multirow}
\usepackage{pdflscape}
\usepackage{setspace}

\title[soft intermediate state of MAXI J1535--571]{Studies on the soft intermediate state X-ray flare of MAXI J1535--571 during its 2017 outburst}

\author[R. C. Ma et al.]{
Ruican Ma,$^{1,2}$
Lian Tao,$^{1}$\thanks{E-mail: taolian@ihep.ac.cn (IHEP)}
Mariano M\'endez,$^{3}$
Shuang-Nan Zhang,$^{1,4}$
Yanjun Xu,$^{1}$
Liang Zhang,$^{1}$
Hexin Liu,$^{1,2}$
\newauthor
Jinlu Qu,$^{1}$
Liming Song,$^{1}$\thanks{E-mail: songlm@ihep.ac.cn (IHEP)},
Xiaoqin Ren,$^{5}$
Shujie Zhao,$^{1,4}$
Yue Huang,$^{1}$
Xiang Ma,$^{1}$
Qingchang Zhao,$^{1,4}$
\newauthor
Yingchen Xu,$^{1,4}$
Panping Li,$^{1,4}$
Zixu Yang,$^{6}$
Wei Yu$^{1,4}$
\\
$^{1}$Key Laboratory of Particle Astrophysics, Institute of High Energy Physics, Chinese Academy of Sciences, Beijing 100049, China\\
$^{2}$Dongguan Neutron Science Center, 1 Zhongziyuan Road, Dongguan 523808, China\\
$^{3}$Kapteyn Astronomical Institute, University of Groningen, P.O. BOX 800, 9700 AV Groningen, The Netherlands\\
$^{4}$University of Chinese Academy of Sciences, Chinese Academy of Sciences, Beijing 100049, China\\
$^{5}$School of Physical and Electronic Sciences, Qiannan Normal University for Nationalities, Duyun 558000, China\\
$^{6}$School of Physics and Optoelectronic Engineering, Shandong University of Technology, Zibo 255000, China
}

\date{Accepted XXX. Received YYY; in original form ZZZ}

\pubyear{2022}

\begin{document}
\label{firstpage}
\pagerange{\pageref{firstpage}--\pageref{lastpage}}
\maketitle

\begin{abstract}
We analyzed an observation with the \textit{Nuclear Spectroscopic Telescope Array} of the black-hole X-ray binary MAXI J1535--571 in the soft intermediate state, in which we detected a 2.5-ks long flare. Our spectral fitting results suggest that MAXI J1535--571 possesses a high spin of $0.97_{-0.10}^{+0.02}$ and a low inclination of approximately $24^{\circ}$. We observed a gradual increase in the inner disc radius, as determined from fits to the continuum spectrum. This trend is inconsistent with an increased flux ratio of the thermal component, as well as the source evolving towards the soft state. This inconsistency may be attributed to a gradual decrease of the color correction factor. Additionally, with a flare velocity of approximately 0.5\,c and a higher hardness ratio during the flare period, the quasi-simultaneous detection of a type-B QPO in the \textit{Neutron Star Interior Composition Explorer} data, and quasi-simultaneous ejecta launch through radio observations collectively provide strong evidence supporting the possibility that the flare originated from a discrete jet ejection.
\end{abstract} 




\section{Introduction}
Black Hole X-ray Binaries (BHXBs) are composed of a black-hole primary and a companion star. The black hole accretes material from the companion star, forming a geometrically thin and optically thick accretion disc around the black hole \citep{Shakura1973}. Most BHXBs are transient systems; after a prolonged period in a quiescent state, they undergo an outburst, likely due to the disc instability mechanism \citep{Tanaka1996,Lasota2001}.

In the X-ray band, the outburst of BHXBs is divided into different states according to their spectral and timing properties \citep[e.g., ][]{Mendez1997,Belloni2002,Remillard2006}. In the early stage of the outburst, BHXBs exhibit a low luminosity, with a non-thermal hard component dominating the spectrum, in the so-called low-hard state (LHS). Typically, in this state, the disc is truncated far away from the innermost stable circular orbit \citep[ISCO;][]{Esin1997, McClintock2001, Done2007}. However, recent reports show that the disc can be at the ISCO even in the LHS \citep[e.g.,][]{Miller2006, Kara2019, Zhang2022}. Additionally, a strong quasi-periodic oscillation (QPO) and a strong band-limited noise component can be detected in the power spectrum in this state \citep[e.g., see the review of][]{Lewin2006}. When the accretion rate of BHXBs continues to increase, they transition into an intermediate state (IMS), with an energy spectrum composed of a comparable non-thermal hard and thermal soft component. The IMS is further divided into the hard-intermediate state (HIMS) and the soft-intermediate state \citep[SIMS;][]{Homan2005}. In the SIMS, the strong QPO and band-limited noise component observed in the HIMS are replaced by a weaker QPO accompanied by weak power-law noise \citep{Belloni2010}. When the thermal soft component dominates the energy spectrum of the BHXBs, the source enters the high soft state (HSS), and the disc is stable and extends down to the ISCO \citep{McClintock2006}. As the outburst evolves, the source returns to the intermediate state and hard state before eventually returning to the quiescent state. A typical outburst follows a counterclockwise ``q'' shape in the hardness intensity diagram \citep[HID;][]{Homan2001, Belloni2005}.

In addition to accreting material from the companion star, BHXBs also exert feedback on their surrounding environment through jets or outflows \citep{Fender2010}. Studying the feedback phenomenon in BHXBs offers the opportunity to understand the characteristics of Active Galactic Nucleus jets and outflows, where feedback can impact the gas dynamics on galactic scales \citep[e.g.,][]{Fabian2012, King2013}. The jets can be divided into two types: i) A persistent, compact, jet appearing in the LHS, for which the radio spectrum is either flat or slightly inverted \citep[$S_{v}\propto$ $\textit{v}^{\alpha}$, with $\alpha \approx 0$; where $S_{v}$ and $\textit{v}$ are radio flux and frequency, respectively;][]{Corbel2000, Fender2001}. ii) A discrete jet, which appears during the intermediate state and exhibits a steep ($\alpha \approx 1$) optically thin spectrum in the radio band \citep[][]{Corbel2004, Lewin2006}. The jet is typically suppressed during the soft state \citep{Coriat2011}.\par

The disc wind is another type of feedback that typically occurs during the HSS and IMS, where the source luminosity is high. The main driving mechanism that is commonly proposed to explain the disc wind \citep{Done2007, Everett2007} involves radiation pressure, which becomes significant when the luminosity of BHXBs approaches the Eddington luminosity. The velocity of the wind is correlated with the luminosity \citep{Matzeu2017}. The two other possible driving mechanisms are thermal \citep[e.g.,][]{Begelman1983,Ponti2012} and magnetic \citep{Blandford1982, Uchida1985}, respectively. Distinguishing between feedback from discrete jets and disc winds can be particularly challenging in high-luminosity BHXBs when both phenomena are present during the IMS. An opportunity to address this arose during the 2017 outburst of the X-ray binary MAXI J1535--571, when a flare occurred in the IMS. \par

MAXI J1535--571 is a black hole candidate \citep{Russell2017}, which was discovered simultaneously with \textit{MAXI}/GSC \citep{Negoro2017} and \textit{Swift}/BAT \citep{Kennea2017} on September 2, 2017. From the analysis of the H\,{\sc i} absorption spectrum from gas clouds along the line of sight the estimated distance of the source is $4.1^{+0.6}_{-0.5}$~kpc  \citep{Chauhan2019}. Using broadband data  from \textit{AstroSat} and the relativistic thin disc model (\texttt{Kerrbb}) during the HIMS, \citet{Sridhar2019} estimate that the mass and distance of the source are $10.4\pm0.6\,{\rm M_{\odot}}$ and $5.4^{+1.8}_{-1.1}$\,kpc, respectively. \citet{Sreehari2019} analyzed \textit{AstroSat} data and applied a two-component accretion flow model, suggesting a mass range of 5--8~${\rm M_{\odot}}$. \citet{Xu2018} analyzed \textit{NuSTAR} data and found a strong reflection component in the LHS. They used the \texttt{relxilllp} and \texttt{relxillpCp} models to limit the spin and inclination of the source, reporting a lower limit on the spin of $a\geq0.84$ with an inclination of ${{57^{+1}_{-2}}}^{\circ}$, and a lower limit on the spin of $a\geq0.987$ with an inclination of ${{75^{+2}_{-4}}}^{\circ}$, respectively. Additionally, they did not find significant evidence of disc truncation in MAXI J1535--571. \citet{Rawat2023b} reported a spin of $0.995\pm{0.001}$, determined by analyzing the type-C QPO using data from \textit{NICER}, \textit{NuSTAR}, and \textit{Swift}/XRT. In the IMS, \citet{Miller2018} used \textit{NICER} data and estimated that the source has a maximal spin of $a=0.994$ and an inclination of $67.4^{\circ}$. They also observed a narrow Fe K emission line, which they attribute to the outer disc with an inclination ${{37^{+22}_{-13}}}^{\circ}$, and they suggest the source has a warped disc. In the SIMS, the inclination of the source was determined to be $\leq45^{\circ}$ based on the characteristics of the jet knots \citep{Russell2019}. The authors attribute the discrepancy with the inclination measured by \citet{Miller2018} to rapid evolution in the disc and jet, or the misalignment between the outer disc and jet. 

\citet{Huang2018}  detected type-C QPOs up to 100 keV using the broadband \textit{Insight}-HXMT \citep{Zhang2014} data of MAXI J1535--571. Subsequently, \citet{Zhang2022b} and \citet{Rawat2023} successfully explained these type-C QPOs by employing the time-dependent Comptonization model \citep{Karpouzas2020, Bellavita2022}, {\tt vkompth}, by making use of \textit{Insight}-HXMT and \textit{NICER} data, respectively. \citet{Stevens2018} reported a weak type-B QPO observed in the SIMS with \textit{NICER} data. The authors noted that the soft lag of this QPO increased as the energy decreased. Furthermore, \citet{Zhang2023} conducted a detailed spectral-timing analysis during the SIMS using \textit{NICER} data. The authors applied the {\tt vkompth} model to constrain the geometry of the corona and proposed a connection between the corona geometry and radio flux as the source transitioned from the HIMS to the SIMS.

In this work we analyze data from MAXI J1535--571 during its SIMS \citep[][]{Tao2018}. We observed a flare using \textit{NuSTAR} (see Section~\ref{sec:lc_hr} below). Given that the source is in the SIMS and exhibits a high luminosity ($\sim4$\,Crab in the 2--20 keV), two potential explanations for this flare arise: discrete jet emission or wind outflow. Our aim is to investigate the physical origin of the flare and to capture the changes in the spectral parameters and components of the source during both the flare and non-flare periods. Furthermore, we also give constraints on the basic parameters of MAXI J1535--571.

In Section~\ref{sec:obs} we introduce the observation information and processing procedure of \textit{NuSTAR} and \textit{NICER} data. In Section~\ref{sec:res}, we present the light curve, hardness ratio (HR), spectral, and timing analysis results. Section~\ref{sec:dis} delves into the system parameters of the source and origin of the flare. The conclusions are given in Section~\ref{sec:con}.

\section{Observations and Data reduction}
\label{sec:obs}

\subsection{NuSTAR}

\textit{NuSTAR} \citep[][]{Harrison2013} has two identical conical Wolter-I approximation optics, paired with the Focal Plane Module (FPM), FPMA, and FPMB, respectively. The instrument covers the energy range from 3 to 79 keV. The observation of MAXI J1535--571 with \textit{NuSTAR} used in this work corresponds to ObsID 80402302006 (MJD 58023.12). The observation log is provided in Tab.~\ref{tab:obs_info}. 

We processed the data with the \textit{NuSTAR} data analysis software (NUSATRDAS) version 1.8.0, using the CALDB version v20180312. Data preprocessing was performed using \texttt{nupipeline}. To remove possible background flares caused by enhanced solar activities, we set \texttt{saacalc=2, saamode=strict, tentcale=no} in \texttt{nupipeline}. Given the significant brightness of MAXI J1535--571 during this observation, greatly exceeding 100 cts\,$\rm{s}^{-1}$, we set \texttt{statusexpr=``STATUS==b0000xxx00xxxx000''}\footnote{\url{https://heasarc.gsfc.nasa.gov/docs/nustar/nustar_faq.html}} within the \texttt{nupipeline} for data processing. To extract the source spectrum, a circular region centered on the source with a radius of 160$''$ was chosen. The background region, situated as far from the central source as possible, was selected with a radius of 120$''$. We use \texttt{nuproducts} to extract the source and background spectra, as well as the ARF and RMF files. To ensure that each bin of the spectra includes a minimum of 50 photons, the $grppha$ command was employed.

\subsection{NICER}
\textit{NICER} \citep[][]{Gendreau2016}, on board the International Space Station (ISS), houses the X-ray Timing Instrument (XTI) as its primary scientific instrument. The XTI comprises 56 X-ray concentrator optics (XRC) that span the energy range of 0.2--12 keV, offering a $\sim$100\,ns time resolution. \textit{NICER} conducted observations of MAXI J1535--571 quasi-simultaneous with \textit{NuSTAR}, specifically obsID 1050360119 (MJD 58023.25). The observation log is found in Tab.~\ref{tab:obs_info}. We used \textit{NICER} data analysis software ({\sc NICERDAS}) version 2022-12-16 V010a and CALDB version xti20221001 to reduce data. We reprocessed the data using \textit{nicerl2} with the standard filtering criteria.

\begin{table}
    \begin{center}
    \renewcommand\arraystretch{1.4}
	\caption{Observation log and count rate of \textit{NuSTAR} and \textit{NICER} of MAXI J1535--571.}
	\label{tab:obs_info}
	\resizebox{\linewidth}{!}{
	\begin{tabular}{cccccc} 
		\hline
		Instrument & obsID & & Start time & Exposure & Rate$^{*}$ \\
		 &  &  & (MJD--58023.12) & (s) & (cts~${\rm s}^{-1}$) \\
		\hline
		\textit{NuSTAR}/FPMA & 80402302006 & epoch 1 & 0.00 & 516 & 1827 \\	
		 & & epoch 2 & 0.06 & 1093 & 1621 \\
		 & & epoch 3 & 0.20 & 2246 & 1508 \\
   \hline
		\textit{NICER} & 1050360119 & orbit 1 & 0.13 & 1913 & 5579 \\	
		 & & orbit 2 & 0.45 & 352 & 5437 \\
		 & & orbit 3 & 0.61 & 352 & 5280 \\   
		\hline
	\end{tabular}
 }
	\end{center}
\footnotesize{$^{*}$ The \textit{NuSTAR} and \textit{NICER} rate covers the 3--79\,keV and 0.3--15\,keV energy band, respectively. 
}	
\end{table}

\begin{figure}
\centering
\includegraphics[width=\columnwidth]{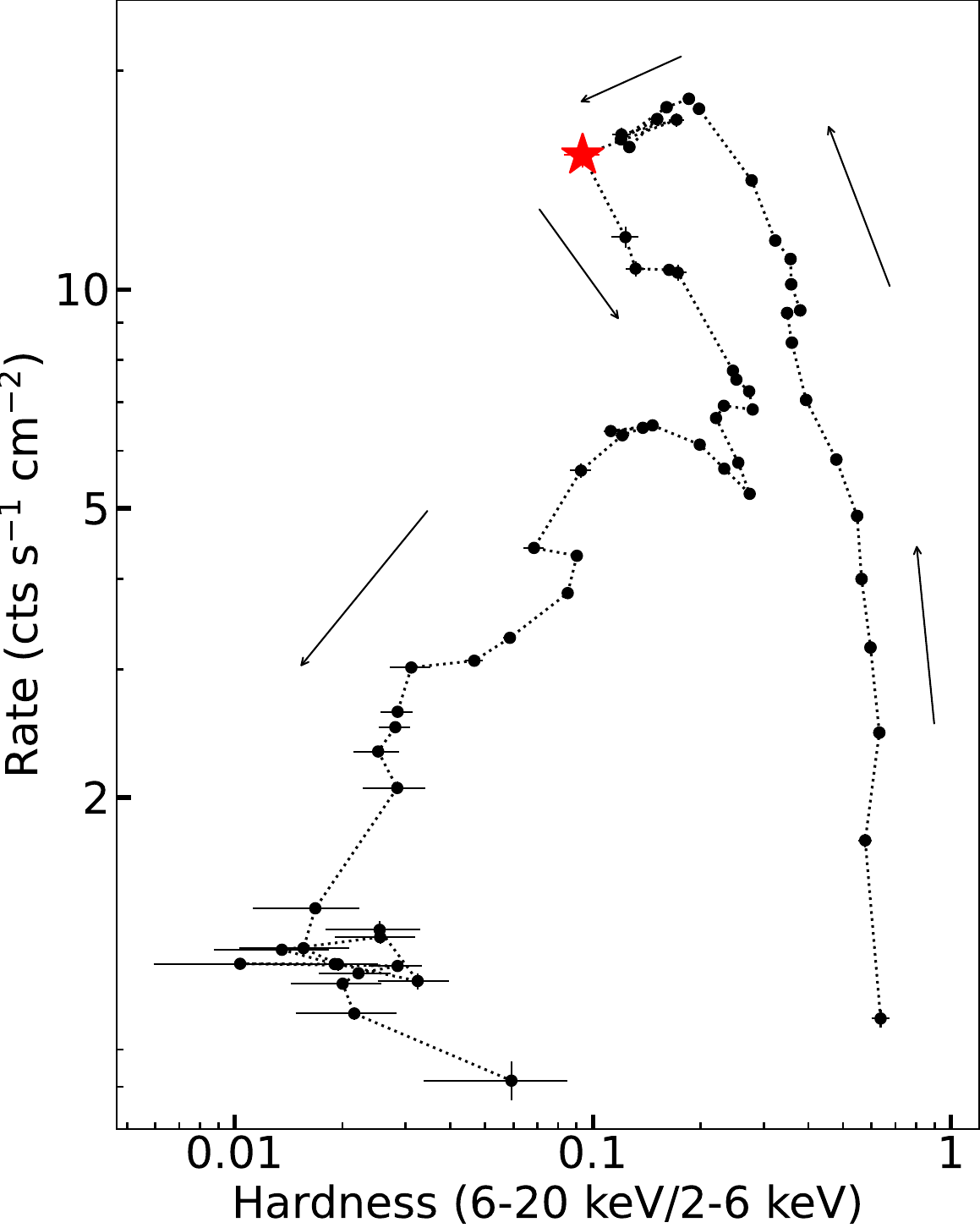} \\
\caption{\textit{MAXI}/GSC HID of MAXI J1535--571 during its 2017 outburst. The intensity is the count rate in the 2--20\,keV, and HR is the ratio of the count rate in the 6--20\,keV to that in the 2--6\,keV. Each point corresponds to one day. The observation used in our works is marked in red. The arrows show the time evolution of the source in this diagram. 
\label{fig:Q}
}
\end{figure}

\begin{figure*}
\centering
\includegraphics[width=0.98\textwidth]{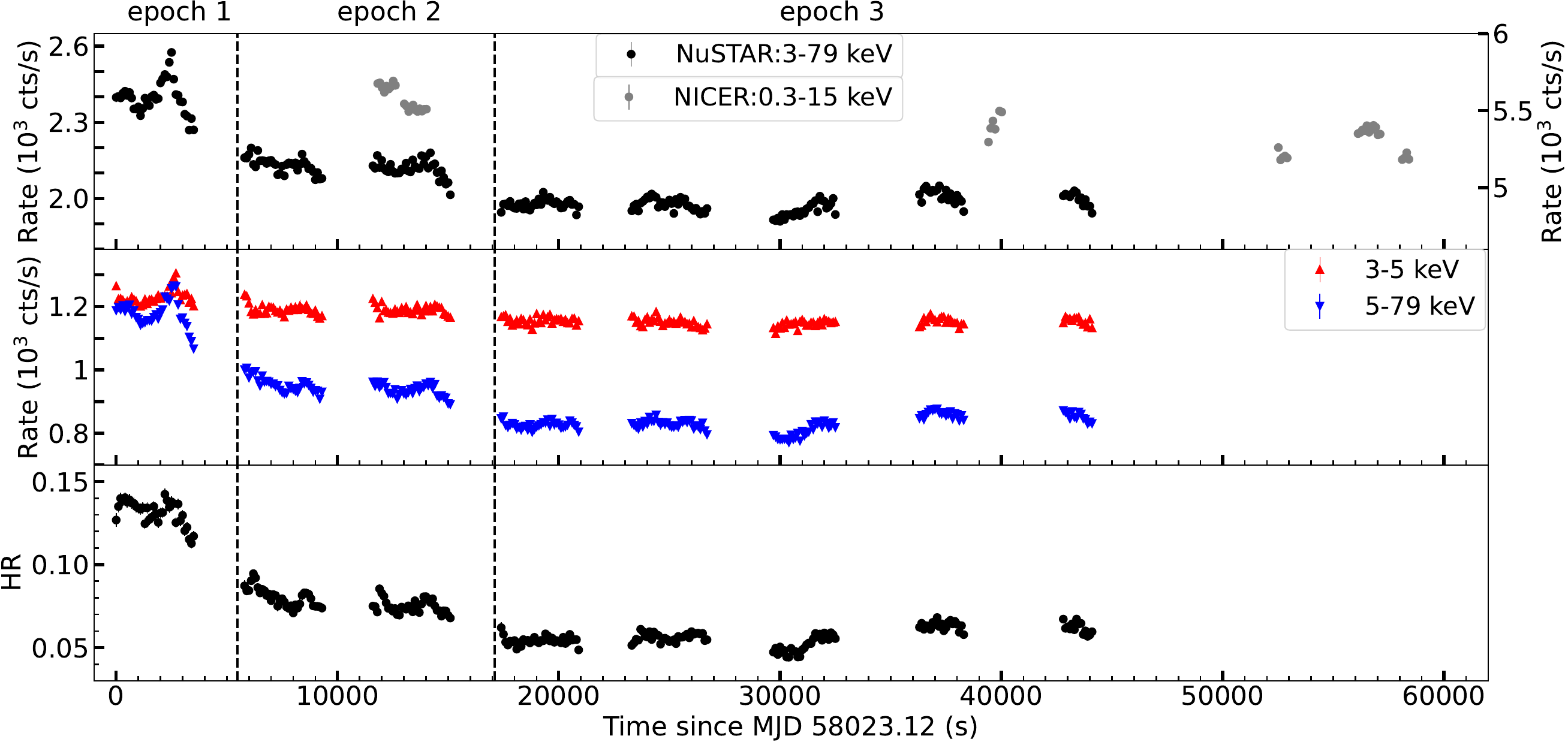} \\
\caption{\textit{NuSTAR} 3--79\,keV (black) and \textit{NICER} 0.3--15\,keV (gray) light curves of MAXI J1535--571 (top panel). \textit{NuSTAR} 3--5\,keV (red) and 5--79\,keV (blue) light curves (middle panel) and HR (bottom panel); the HR defined as the ratio of the count rate in the 10--79\,keV band to that in the 3--5\,keV band. The \textit{NuSTAR} observation is divided into three epochs. A flare is apparent in epoch 1 at $\sim$1\,ks from the start of the observation. Only the \textit{NuSTAR}/FPMA data are shown for clarity. The time resolution in all panels is 100\,s.
\label{fig:LC_HR}
}
\end{figure*}

\section{Results}
\label{sec:res}

\subsection{Hardness Intensity Diagram}
\label{sec:hid}

For a comprehensive overview of the outburst of MAXI J1535--571, we present an HID utilizing MAXI/GSC data\footnote{\url{http://maxi.riken.jp/mxondem/}} spanning from MJD 58000 to MJD 58168 (Fig.~\ref{fig:Q}). The intensity is the count rate within the energy range of 2--20\,keV, and the HR is defined as the ratio of the count rate in the hard energy band (6--20\,keV) to that in the soft energy band (2--6\,keV). Each data point corresponds to a single day. The HID of the whole 2017 outburst exhibits a ``q'' shape commonly observed in classical outbursts of black hole transients \citep[e.g.,][]{Homan2001,Belloni2016}. Our study specifically focuses on analyzing the data corresponding to the high-brightness soft intermediate state, denoted by the red star marker.

\subsection{Light Curve and Hardness Ratio}
\label{sec:lc_hr}

As shown in the top panel of Fig.~\ref{fig:LC_HR}, the \textit{NuSTAR} light curve is divided into three epochs based on the count rate: epoch 1, epoch 2, and epoch 3, with a flare occurring during epoch 1 (see also Tab.~\ref{tab:obs_info}). In epoch 1: At about 1\,ks from the start of the observation, the count rate rapidly increases from $2.3\times10^{3}~{\rm cts~s^{-1}}$ to $2.6\times10^{3}~{\rm cts~s^{-1}}$ in about 1.5\,ks, and then returns to $2.3\times10^{3}~{\rm cts~s^{-1}}$ in about 1\,ks. The whole flare lasts for about 2.5\,ks. After that, in epoch 2, the count rate gradually decreases from $2.2\times10^{3}~{\rm cts~s^{-1}}$ to $2.0\times10^{3}~{\rm cts~s^{-1}}$, and remains more or less constant at that value in epoch 3. 

The \textit{NICER} observation, consisting of a total of 3 orbits (see Tab.~\ref{tab:obs_info}), unfortunately, missed the flare event but captured the post-flare period. In this period, the source count rate exhibited a gradual decline from $\sim 5.6\times10^{3}~{\rm cts~s^{-1}}$ to $\sim 5.2\times10^{3}~{\rm cts~s^{-1}}$ (top panel of Fig.~\ref{fig:LC_HR}).

We present the \textit{NuSTAR} light curve in two energy bands in the middle panel of Fig.~\ref{fig:LC_HR}. In epoch 1, the count rate of the soft energy band (3--5\,keV) remains more or less constant at $1.2\times10^{3}~{\rm cts~s^{-1}}$, with a minor upward trend to $1.3\times10^{3}~{\rm cts~s^{-1}}$, whereas the count rate in the hard energy band (5--79\,keV) experiences a significant increase, from $1.0\times10^{3}~{\rm cts~s^{-1}}$ to $1.3\times10^{3}~{\rm cts~s^{-1}}$. In epoch 2 and epoch 3, the count rate in the soft energy band decreases slightly from $1.3\times10^{3}~{\rm cts~s^{-1}}$ to $1.1\times10^{3}~{\rm cts~s^{-1}}$, whereas in the hard band the count rate rapidly decreases from $1.3\times10^{3}~{\rm cts~s^{-1}}$ to $0.8\times10^{3}~{\rm cts~s^{-1}}$. 

The HR, calculated as the ratio of photons in the 10--79 keV energy band to those in the 3--5 keV, is shown in the bottom panel of Fig.~\ref{fig:LC_HR}. In epoch 1 the source is hard, HR=0.15, whereas, in epoch 2 and epoch 3, HR decreases from 0.09 to 0.05.

\begin{figure*}
\centering
\includegraphics[width=0.98\textwidth]{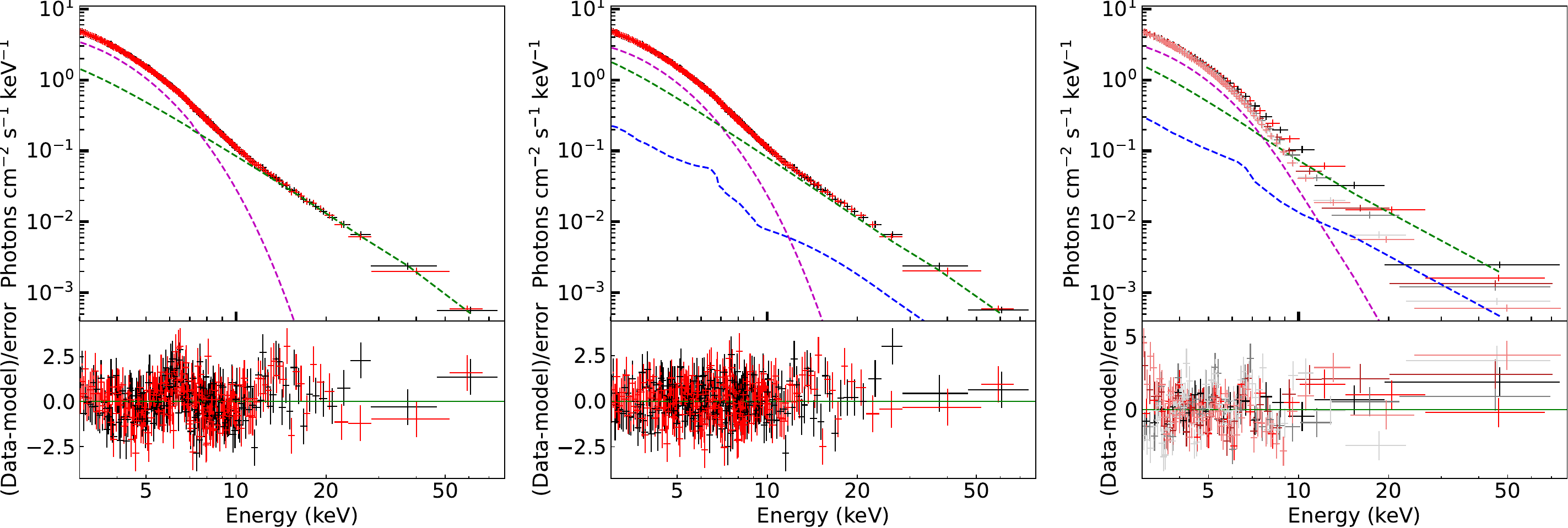} \\
\caption{\textit{NuSTAR} energy spectra of MAXI J1535--571 in the 3--79\,keV band. Left panel: The representative spectra (epoch 1) are fitted using the model \texttt{tbabs*(diskbb+powerlaw)} (top), along with the residuals of the best-fitting model (bottom). Each spectrum of the three epochs is individually fitted. Middle panel: Similar to the left panel but using the \texttt{tbabs*(diskbb+cutoffpl+relxilllp)} model. Right panel: The spectra of the three epochs are fitted jointly, using the model \texttt{tbabs*(diskbb+cutoffpl+relxilllp)}. The data from FPMA (FPMB) for epochs 1 to 3 are denoted by black (red), gray (dark red), and light gray (light red), respectively. The magenta, green, and blue dotted dashed lines represent the {\tt diskbb}, {\tt cutoffpl} and {\tt relxilllp} components, respectively. For plotting purposes, we rebin the data.
\label{fig:pha}
}
\end{figure*}

\begin{figure}
\centering
\includegraphics[width=\columnwidth]{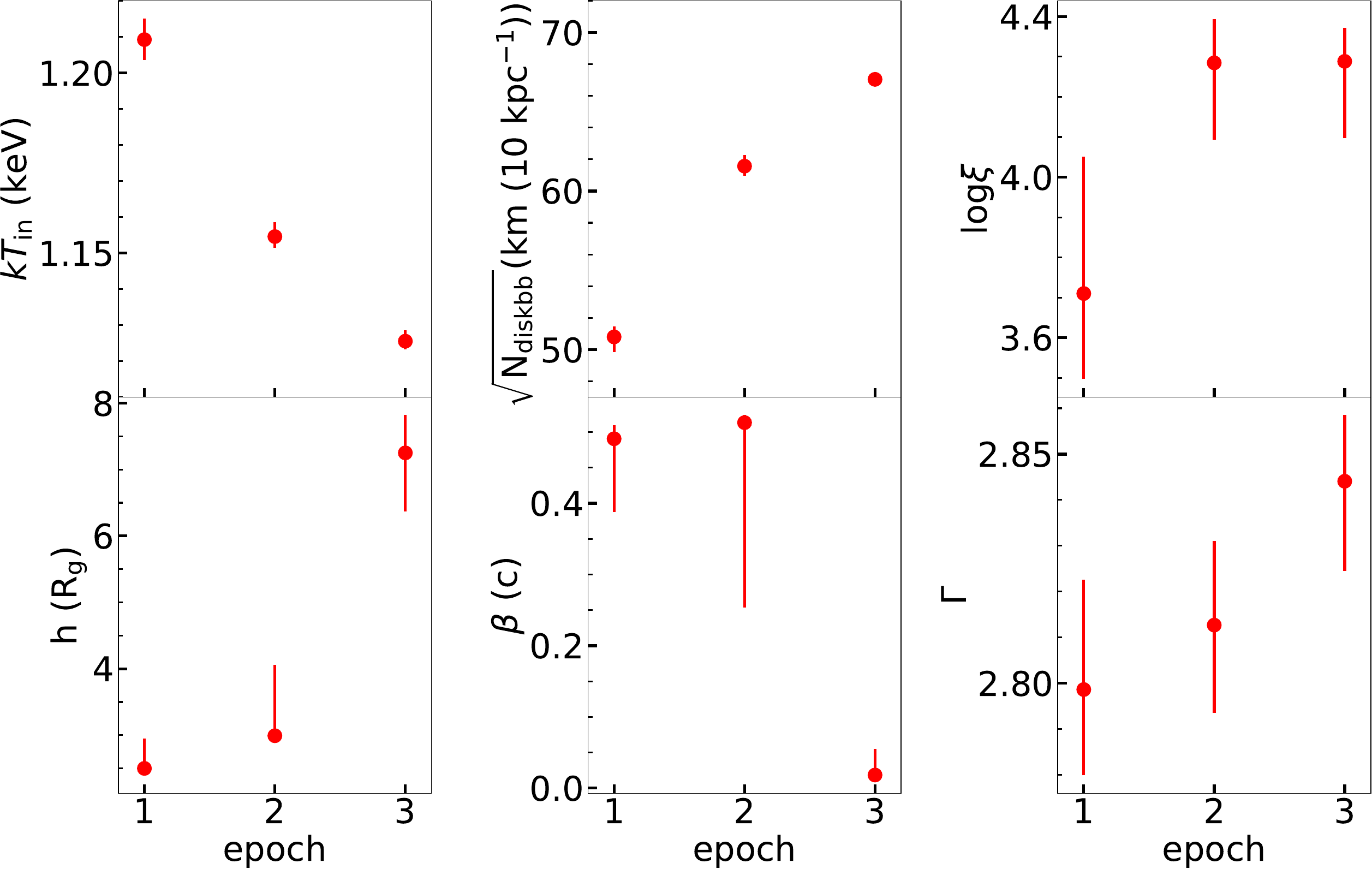} \\ 
\caption{Evolution of the best-fitting parameters for the joint fitting of MAXI J1535--571 using the \texttt{tbabs*(diskbb+cutoffpl+relxilllp)} model. The parameter $kT_{\rm in}$ is the disc temperature, $\sqrt{N_{\rm diskbb}}$ is the square root of {\tt diskbb} normalization (in units of ${\rm km\,(10\,kpc)^{-1}}$), log\,$\xi$ is the ionization parameter of the accretion disc, $h$ is the corona height in units of $R_{\rm g}$, $\beta$ is the velocity of the primary source in units of $c$, and $\Gamma$ is the photon index.
\label{fig:pars}
}

\end{figure}

\begin{table}
    \begin{center}
    \large
    \renewcommand\arraystretch{1.7}
	\caption{Parameters of the best-fitting spectral model, {\tt tbabs*(diskbb+cutoffpl+relxilllp)}, in MAXI J1535--571 from individual fittings for the three epochs.}
	\label{tab:pars}
	\resizebox{\columnwidth}{!}{
	\begin{tabular}{lcccc} 
		\hline
		Component & Parameter & {epoch 1} & {epoch 2} & {epoch 3} \\
		\hline
		{\tt tbabs} & $N_{\rm H}\,(10^{22}\,\rm{cm^{-2}})$ & $3.6\pm{0.3}$ & $3.2\pm{0.2}$ & $3.03_{-0.10}^{+0.14}$ \\          
		{\tt diskbb} & $kT_{\rm in}$\,(keV) & $1.186_{-0.017}^{+0.012}$ & $1.153\pm{0.007}$ & $1.130\pm{0.005}$ \\   
		& $\sqrt{N_{\rm diskbb}}\,{\rm (km\,(10\,kpc^{-1}))}$  & $52.8_{-1.3}^{+1.6}$ & $61.9\pm{1.1}$ & $66.9_{-0.6}^{+1.0}$ \\   
		{\tt cutoffpl} & $\Gamma$ & $2.84\pm{0.05}$ & $2.82\pm{0.04}$ & $2.82\pm{0.05}$  \\
		& $N_{\rm cpl}$ & $59_{-7}^{+9}$ & $22\pm{5}$ & $13\pm{4}$ \\
		{\tt relxilllp} & $h$\,($R_{\rm g}$) & $5_{-2}^{+6}$ & $3.0_{-0.8}^{+2.1}$ & $6\pm{2}$ \\   
		& $\beta$\,($c$) & $0.6\pm{0.3}$ & $0.5\pm{0.4}$ & $0.08_{-0.07}^{+0.26}$ \\ 
		& $a$ & $0.99_{-0.83}^{*}$ & $0.972_{-0.470}^{+0.007}$ & $0.970_{-0.726}^{+0.014}$ \\ 
		& $i$\,($^{\circ}$) & $20_{-13}^{+8}$ & $26\pm{5}$ & $21\pm{5}$ \\ 
		& $R_{\rm in, lp}$ (ISCO) & $1^{\dagger}$ & $1^{\dagger}$ & $1^{\dagger}$ \\ 
		& ${\rm log}\,\xi\,(\rm{log[erg\,cm\,s^{-1}}])$ & $3.9\pm{0.4}$ & $4.27_{-0.26}^{+0.10}$ & $4.29_{-0.19}^{+0.14}$  \\ 
	  & $A_{\rm Fe}$ (solar) & $0.9_{-0.3}^{+0.9}$ & $0.8\pm{0.3}$ & $0.7\pm{0.3}$ \\  
		& $N_{\rm ref}$ & $1.0_{-0.5}^{+0.9}$ & $0.9_{-0.4}^{+1.1}$ & $0.35_{-0.05}^{+0.45}$ \\ 
        \hline      
        & $\chi^{2}/v$ & 1059.70/1052 & 1169.42/1088 & 1376.35/1183 \\
        \hline
	\end{tabular}
	}
	\end{center}
\footnotesize{$^{*}$ The positive error of the parameter is pegged at the upper limit. \\
$^{\dagger} R_{\rm in}$ in \textit{relxilllp} component is fixed at the ISCO. \\
}	
\end{table}

\begin{table}
  \begin{center}
  \renewcommand\arraystretch{1.7}
	\caption{Parameters of the best-fitting spectral model, {\tt tbabs*(diskbb+cutoffpl+relxilllp)}, in MAXI J1535--571 from joint fittings for the three epochs.}
	\label{tab:jointfit_pars}
	\resizebox{\columnwidth}{!}{
	\begin{tabular}{lcccc} 
		\hline
		Component & Parameter & {epoch 1} & {epoch 2} & {epoch 3} \\
		\hline
		{\tt tbabs} & ${N_{\rm H}}^{*}\,(10^{22}\,\rm{cm^{-2}})$ &  & $3.14\pm{0.08}$ & \\     
		{\tt diskbb} & $kT_{\rm in}$\,(keV) & $1.209\pm{0.006}$ & $1.155\pm{0.004}$ & $1.126\pm{0.003}$ \\  
		& $\sqrt{N_{\rm diskbb}}\,{\rm (km\,(10\,kpc^{-1}))}$ & $50.8_{-1.0}^{+0.7}$ & $61.6\pm{0.7}$ & $67.0\pm{0.5}$ \\  
		{\tt cutoffpl} & $\Gamma$ & $2.80\pm{0.02}$ & $2.81\pm{0.02}$ & $2.84_{-0.02}^{+0.01}$ \\
		& $N_{\rm cpl}$ & $48\pm{3}$ & $22.4_{-0.9}^{+1.3}$ & $14.4\pm{0.7}$  \\
		{\tt relxilllp} & $h$\,($R_{\rm g}$) & $2.5_{-0.1}^{+0.5}$ & $2.99_{-0.02}^{+1.07}$ & $7.3_{-0.9}^{+0.6}$  \\  
		& $\beta$\,($c$) & $0.49_{-0.10}^{+0.02}$ & $0.51_{-0.26}^{+0.01}$ & $0.0184_{-0.0005}^{+0.0363}$ \\ 
		& $a^{*}$ &  & $0.97_{-0.10}^{+0.02}$ &   \\ 
		& $i^{*}$\,($^{\circ}$) &  & $24\pm{2}$ &  \\ 
		& ${R_{\rm in, lp}}^{*}$ (ISCO) &  & $1^{\dagger}$ &  \\ 
		& ${\rm log}\,\xi\,(\rm{log[erg\,cm\,s^{-1}}])$ & $3.7\pm{0.3}$ & $4.28_{-0.19}^{+0.11}$ & $4.29_{-0.19}^{+0.08}$  \\ 
	 & ${A_{\rm Fe}}^{*}$ (solar) &  & $0.80\pm{0.07}$ & \\ 
		& $N_{\rm ref}$ & $1.384\pm{0.002}$ & $0.995_{-0.397}^{+0.009}$ & $0.30_{-0.02}^{+0.05}$  \\ 
    \hline
    & $\chi^{2}/v$ &  & 3613.27/3331 &  \\
    \hline
	\end{tabular}
	}
	\end{center}
\footnotesize{$^{*}$ Parameters linked together during joint fitting.\\
$^{\dagger} R_{\rm in}$ in \textit{relxilllp} component is fixed at ISCO. \\
}	
\end{table}

\subsection{Spectral Analysis}
\label{sec:spec}

We analyzed the energy spectra using XSPEC v12.13.0c \citep{Arnaud1996}. Since \textit{NICER} did not capture the flare, we restricted our spectral analysis to the 3--79 keV \textit{NuSTAR} data. In our fitting process, we use {\tt tbabs} to account for neutral absorption from the interstellar medium in the direction of the source. The abundance and cross-section tables employed are from \citet{Wilms2000} and \citet{Verner1996}, respectively. We introduce an energy-independent multiplicative factor that was allowed to vary freely for FPMB but kept fixed at 1 for FPMA. We use the Goodman–Weare algorithm \citep{Goodman2010} in a Monte Carlo Markov Chain (MCMC) to compute parameter errors. The chain employs 160 walkers and generates a total of 160,000 samples utilizing a Cauchy proposal distribution. To ensure convergence, the initial 320,000 steps of the MCMC are discarded. The uncertainties of each parameter correspond to the 90 per cent confidence range.

\subsubsection{Fitting spectra individually}
\label{res:fit_ind}
First, we individually fit the spectra of each epoch using an absorbed multi-color disc and a cutoff power-law component, {\tt tbabs*(diskbb+cutoffpl)} in XSPEC notation. As shown in the left panel of Fig.~\ref{fig:pha}, the fit with this model shows significant residuals at 6--7\,keV and at around 15\,keV, which are likely due to a reflection component (note that here we give the spectral fitting results of epoch 1; epoch 2 and epoch 3 show similar residuals). 

Therefore, we add a relativistic reflection component, denoted as {\tt relxilllp}, from the {\tt relxill} model family \citep[version 2.3;][]{Dauser2014,Garcia2014}. The {\tt relxilllp} component assumes a lamppost geometry, where the corona is a point illuminating source situated at a height $h$ above the accretion disc. Although recent findings, based on polarization results, suggest a potentially horizontally extended corona in both the LHS and HIMS \citep[e.g.,][]{Krawczynski2022, Veledina2023}, there are no polarization measurements available for the SIMS to confirm a similar corona geometry. In a study by \citet{Ma2023}, who utilized \textit{NICER} data to investigate the evolution of the accretion geometry of MAXI J1820+070 from HIMS to SIMS during the 2018 outburst. The authors reported a horizontally extended corona in the HIMS (supported by polarization results) and a vertically extended corona in the SIMS. Importantly, the {\tt relxilllp} model provides the velocity of the illuminating source, introducing an additional constraint on the source properties. Therefore, in our analysis, we choose the lamp-post model, \texttt{relxilllp}, for spectral fitting. 

The current model is {\tt tbabs*(diskbb+cutoffpl+relxilllp)}. We set $switch\_ returnrad=1$ to consider the return radiation, $switch\_reflfrac\_boost=1$ to use the $boost$ parameter in our fitting and set $boost=-1$ to get only the reflection emission from {\tt relxilllp}\footnote{\url{http://www.sternwarte.uni-erlangen.de/~dauser/research/relxill/}}. We link $\Gamma$ and $E_{\rm cut}$ in {\tt relxilllp} with the corresponding parameters $\Gamma$ and $E_{\rm cut}$ in {\tt cutoffpl}. We initially allowed the disc inner radius, $R_{\rm in}$, to fit freely, but it tended to converge towards the ISCO. Therefore, we decided to fix $R_{\rm in}$ at the ISCO \citep[see also][]{Dong2022} and treated the spin as a free-fitting parameter. It is important to note that the {\tt relxilllp} component not only gives the spin, inclination, iron abundance, and corona height but also yields the velocity of the illuminating source. After considering the reflection component, the reduced chi-square decreases from 1163.26 for 1059 degrees of freedom to 1059.70 for 1052 degrees of freedom. This fit is in the middle panel of Fig.~\ref{fig:pha}. 

The evolution of the parameters is presented in Tab.~\ref{tab:pars}, and the representative corner plot is shown in Fig.~\ref{fig:Cont_epoch1} to illustrate the degeneracy between different parameters. In all epochs the hydrogen column density, and inclination remain more or less constant, $N_{\rm H}\sim3.2\times10^{22}~{\rm cm^{-2}}$ and $i\sim23^{\circ}$, respectively. The spin of MAXI J1535--571 cannot be constrained very well, with $a>0.16$. The iron abundance is about 0.8 times solar, and the ionization parameter of the accretion disc is ${\rm log}\,\xi \sim$ 3.9--4.3 with large error bars. 

From epoch 1 to epoch 3, the disc temperature decreases from $\sim$1.2\,keV to $\sim$1.1\,keV. However, there is a peculiar behavior of the apparent disc inner radius, $r_{\rm in}$. It significantly increases from $(53\pm{2})\,\frac{D_{10}}{\sqrt{\cos{i}}}$\,km  to $(67\pm1)\,\frac{D_{10}}{\sqrt{\cos{i}}}$\,km, where $i$ and $D_{\rm 10}$ are the source inclination and distance in units of 10\,kpc, respectively. At the same time, the disc flux fractional contribution to the total flux shows an increasing trend and the source is transiting to the HSS (details see Section~\ref{sec:flux}). 

Throughout this period, the height of the illuminating source remains more or less constant at $\sim5\,R_{\rm g}$. At the same time, the velocity of the illuminating source decreases from approximately 0.6\,c in epoch 1, rapidly approaching zero by epoch 3. The reflection fraction for the three epochs is $\sim$0.59, $\sim$1.23 and $\sim$1.48, respectively. The cutoff energy values are not well constrained, reaching the upper limit; therefore, they are not provided. It is noteworthy that in our individual spectral fitting, certain parameters, particularly spin, prove challenging to constrain effectively. Consequently, we undertake a joint fitting approach for the spectra of these three epochs (see Section~\ref{res:joint_fit}).

\subsubsection{Fitting spectra jointly}
\label{res:joint_fit}

For the joint fitting of the spectra for three epochs, we used the same parameter settings as those used in individual fitting, as described in Section~\ref{res:fit_ind}. Here, however, we linked the hydrogen column density ($N_{\rm H}$), spin ($a$), inclination ($i$), and iron abundance ($A_{\rm Fe}$), across the three epochs. The spectra can be fitted well, as shown in the right panel of Fig.~\ref{fig:pha}. The fitting results are presented in Tab.~\ref{tab:jointfit_pars} and Fig.~\ref{fig:pars}, and the representative corner plot is shown in Fig.~\ref{fig:Cont_joint_epoch1} to illustrate the degeneracy between different parameters. Notably, $N_{\rm H}=(3.14\pm{0.08})\times10^{22}\,{\rm cm^{-2}}$, $a=0.97_{-0.10}^{+0.02}$, $i=24^{\circ}\pm{2}$, and $A_{\rm Fe}=0.80\pm{0.07}$. In comparison to the individual fitting, these parameters exhibit similar fitted values but with smaller error bars. Particularly noteworthy is the improved precision in constraining the spin. The cutoff energy values are still poorly constrained, reaching the upper limit.

The peculiar behavior observed in individual fitting persists in the joint fitting: The apparent inner disc radius increases from $(51\pm{1})\,\frac{D_{10}}{\sqrt{\cos{i}}}$\,km to $(67\pm{1})\,\frac{D_{10}}{\sqrt{\cos{i}}}$\,km corresponding to the increased flux ratio of the disc component to the total component. Simultaneously, the corona height, $h$, exhibits a notable increasing trend, from $\sim2.5\,{R}_{\rm g}$ increases to $\sim7.3\,{R}_{\rm g}$. The $\beta$ decreases from $0.5\,c$ to zero. $\beta$ and $h$ show a slight degenerative trend (see Fig.~\ref{fig:Cont_joint_epoch1}). The reflection fraction increases from $\sim$1.44 to $\sim$1.54. It is evident that both individual and joint fittings yield similar fitted values, yet the latter provides more robust constraints on the parameters. Therefore, in the ensuing discussion in Section~\ref{sec:dis}, we will rely on the results obtained from the joint fitting.

\subsection{Flux Evolution}
\label{sec:flux}

\begin{figure}
\centering
\includegraphics[width=\columnwidth]{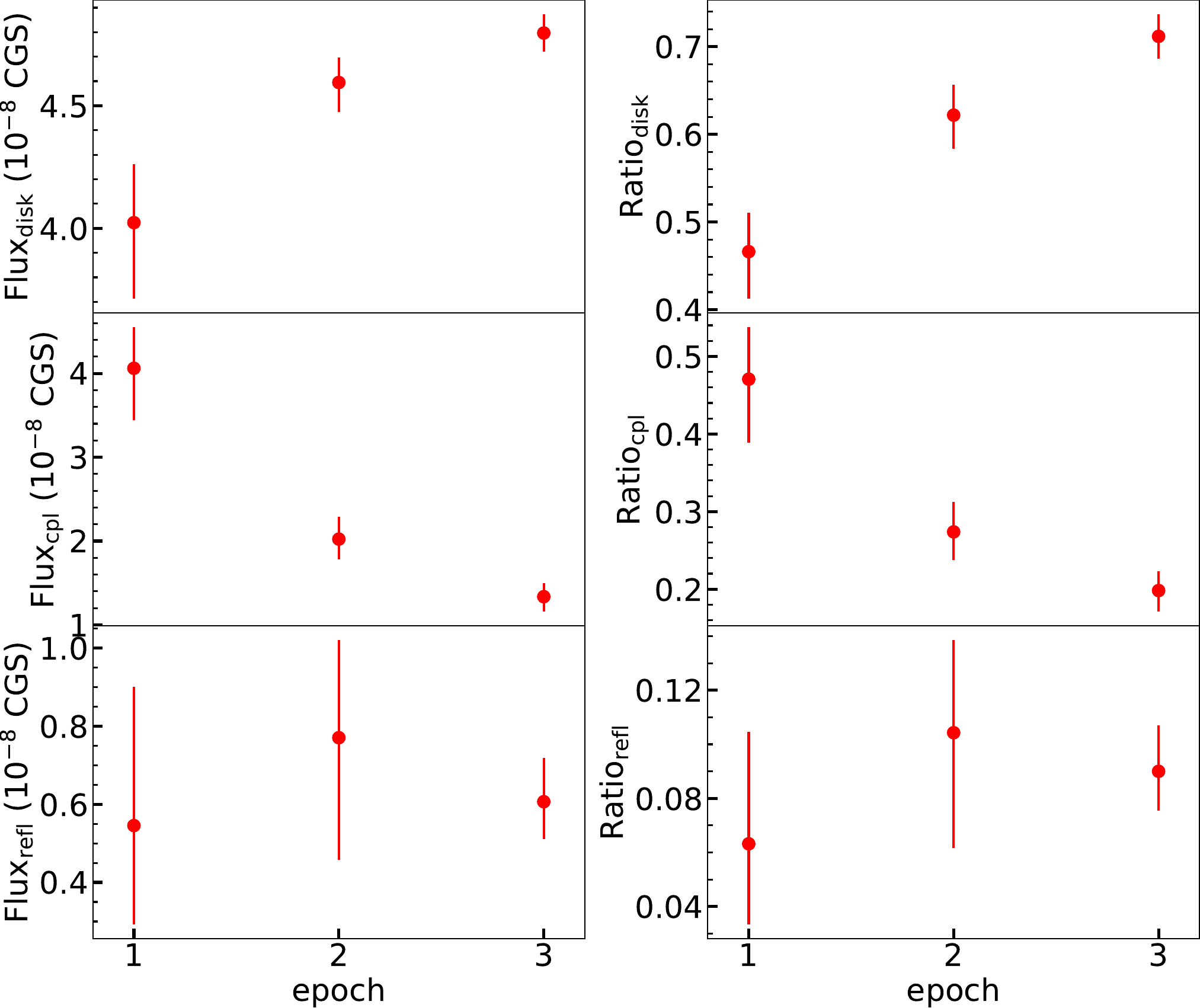} \\ 
\caption{Left panel: From top to bottom, unabsorbed flux of the {\tt diskbb}, {\tt cutoffpl} and reflection components in MAXI J1535--571 in the 3--79\,keV band. Right panel: The ratio of the flux of each component to the total flux.
\label{fig:flux}
}
\end{figure}

The evolution of the unabsorbed 3--79\,keV flux and the fractional contribution of the different components in the three epochs is presented in Fig.~\ref{fig:flux}. From epoch 1 to epoch 3 the {\tt diskbb} flux increases from $4.0\times10^{-8}\,{\rm erg\,cm^{-2}\,s^{-1}}$ to $4.8\times10^{-8}\,{\rm erg\,cm^{-2}\,s^{-1}}$, and its fractional contribution to the total flux increases from 0.46 to 0.71. On the contrary, the {\tt cutoffpl} flux decreases from $4.1\times10^{-8}\,{\rm erg\,cm^{-2}\,s^{-1}}$ to $1.3\times10^{-8}\,{\rm erg\,cm^{-2}\,s^{-1}}$, while its fractional contribution to the total flux decreases from 0.47 to 0.20. The reflection component remains more or less constant at $0.6\times10^{-8}\,{\rm erg\,cm^{-2}\,s^{-1}}$ with large error bars, and its fractional contribution to the total flux remains more or less constant at 0.1.

\begin{figure}
\centering
\includegraphics[width=\columnwidth]{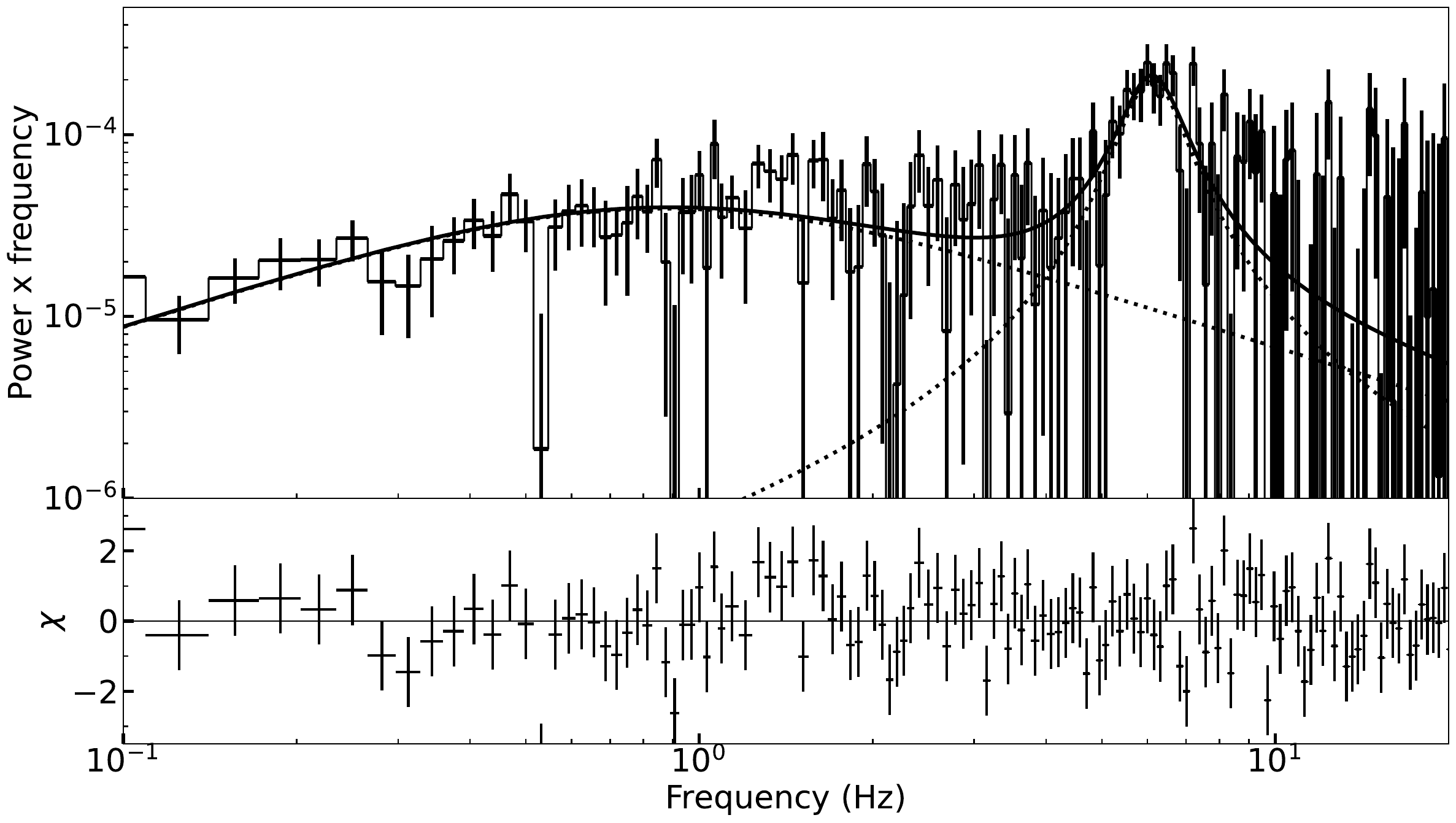} \\ 
\caption{Representative PDS of MAXI J1535--571 in the 0.5–12 keV energy band with \textit{NICER} data. The solid and dashed lines represent the best-fitting total model and multi-Lorentzian function, respectively. 
\label{fig:pds}
}
\end{figure}

\subsection{Timing Properties}
\label{sec:timing}

The dead time of \textit{NuSTAR} of $\sim$2.5\,ms leads to an increase of the noise level at high frequency that is frequency dependent \citep{Bachetti2015}. Therefore, we extract the cospectrum instead of the power density spectrum (PDS) in the full energy band (3--79\,keV), using the tool {\sc hendrics}\footnote{\url{https://hendrics.stingray.science/en/stable/index.html}} based on the {\sc stingray} pacakge\footnote{\url{https://docs.stingray.science/dataexplo.html}} \citep{Huppenkothen2019}. For each epoch, we use a segment length of 32\,s, and a time resolution of 1/128\,s, resulting in a Nyquist frequency of 64\,Hz. The cospectrum is normalized to rms units \citep{Belloni1990}. We apply a logarithmic rebin in frequency, where each bin size is exp(3/100) times larger than the previous one. No obvious QPO signals are detected in any of the three epochs. 

We compute the average PDS for each orbit of \textit{NICER} data using the same parameter settings as \textit{NuSTAR}. This analysis encompasses the energy range from 0.5 keV to 12 keV. We ignore the background count rate when we normalize the power spectrum to rms units due to its insignificance when compared to the source count rate. We detect a weak type-B QPO in the \textit{NICER} data in this whole observation (see Fig.~\ref{fig:pds}). The QPO is detectable in the \textit{NICER} data rather than the \textit{NuSTAR} data, possibly due to the former having a large effective area.

\section{Discussion}
\label{sec:dis}

We conduct an analysis of the observation from \textit{NuSTAR} and its quasi-simultaneous observation from \textit{NICER} for MAXI J1535--571 during the SIMS. Our specific focus lies on a 2.5\,ks X-ray flare that was observed by \textit{NuSTAR}. Our analysis delves into the spectral and timing properties of MAXI J1535--571 during this event and found that the source possesses a high spin ($a=0.97_{-0.10}^{+0.02}$) and a low inclination ($i\sim24^{\circ}$).  Moreover, we explained the origin of the flare.

\subsection{System Parameters}
\label{sec:basic_pars}

The hydrogen column density, $N_{\rm H}$, remains more or less constant across the three epochs, at around $3\times10^{22}\,{\rm cm^{-2}}$. This value is consistent with the result of $\sim(2.4-3.1)\times10^{22}\,{\rm cm^{-2}}$ obtained from fitting the \textit{AstroSat} data \citep{Sridhar2019}. The photon index, $\Gamma$, is about 2.8, which is consistent with the result of $2.6^{+0.4}_{-0.3}$ in \citet{Tao2018}. These results imply that the source has transitioned from the HIMS to the SIMS. 

The spin of MAXI J1535--571, $a=0.97_{-0.10}^{+0.02}$, is consistent with the previous reports, $a>0.84$, $a\sim0.994$ and $a\sim0.985$ obtained by fitting the \textit{NuSTAR} data \citep{Xu2018, Dong2022} and the \textit{NICER} data \citep{Miller2018}, respectively. These results support that the source has an extremely high spin. The iron abundance is about 0.8 times solar, consistent with previous results of $\sim0.62$ \citep{Miller2018} and 0.8--1.4 \citep{Xu2018}. This confirms that the source has a low iron abundance. 

The inclination, $i=24^{\circ}\pm{2}$, is essentially consistent with the results given by \citet{Russell2019}, who limited the inclination of the jet to $<45^{\circ}$ based on the kinematics of the jet knots. By fitting the \textit{NICER} data, the inclination is $i={37^{+22}_{-13}}^{\circ}$ based on the distant reflection, while the broad iron line gives an inclination of $\sim67^{\circ}$; \citet{Miller2018} interpreted this as the source having a warped accretion disc. However, \citet{Xu2018} and \citet{Dong2022} reported a higher inclination for the source, $\sim60^{\circ}$ and $\sim70^{\circ}$, respectively. \citet{Dong2022} interpreted this high inclination as the angular momentum of the disc being misaligned with the radio jet, and the jet itself misaligned with the binary plane. The small values measured in this work may be attributed to the presence of an evolving warped accretion disc. The disc, which is misaligned and distorted, undergoes rapid evolution primarily due to the Lense-Thirring effect \citep[e.g.,][]{Nixon2012, Nixon2014}. 

\subsection{Peculiar Inner Disc Radius Behavior} 
\label{sec:fcolor}

As shown in Fig.~\ref{fig:pars}, the inner disc radius given by the {\tt diskbb} component indicates an increasing trend with an increased disc flux, as shown in the top panel of Fig.~\ref{fig:flux}. Taking into account the physical inner disc radius, $R_{\rm in}$, and the apparent inner disc radius, $r_{\rm in}$, which satisfy the relationship \citep{Kubota1998}:
\begin{equation}
\label{equ1}
R_{\rm in}=\xi f_{\rm col}^{2} r_{\rm in} = \xi f_{\rm col}^{2} \sqrt{N_{\rm diskbb}/{\rm cos}i}D_{\rm 10},
\end{equation}
where $\xi \approx 0.415$ represents the geometric correction factor, and $f_{\rm col}$ is the color correction factor \citep{Shimura1995,Zhang1997}. By selecting typical values of $f_{\rm col}\sim1.7$, $i={24\pm{2}}^{\circ}$ from our fitting, $D_{\rm 10}=5.4$\,kpc and $M=10.4\,{\rm M_{\odot}}$ \citep{Sridhar2019}, we can calculate $R_{\rm in}$ to be $34\pm{1}$\,km (2.2\,$R_{\rm g}$), $41\pm{1}$\,km (2.7\,$R_{\rm g}$), and $45\pm{1}$\,km (2.9\,$R_{\rm g}$), respectively, showing a significant increasing trend. 

Given that the source is currently in the SIMS and undergoing a transition to the HSS, and the disc flux fractional contribution to the total flux increases, the inner radius of the disc should not increase. The inconsistency in the inner disc radius could be linked to changes in the color correction factor $f_{\rm col}$. To mitigate the impact of uncertainties stemming from other parameters, we assume that the inner radius of the disc remains stable at the ISCO and compute the ratio of $f_{\rm col}$ for these three epochs, normalized with respect to the first epoch. The resulting ratio is 1:0.91:0.87 (where $\xi$, $i$ and $D_{\rm 10}$ remain unaltered), indicating a declining trend of $f_{\rm col}$. This implies that the peculiar behavior of the inner disc radius, where the inner disc radius given by the {\tt diskbb} component increases with increasing disc flux, should be attributed to a decrease of $f_{\rm col}$.

The evolution of the color correction $f_{\rm col}$ has been previously studied. When the source is in a disc-dominated state, $f_{\rm col}$ has been found to be proportional to the accretion rate $\dot{M}$ \citep{Done2008}. Recent research on the outburst evolution of MAXI J1348--630 by \citet{Zhang2022} reveals that $f_{\rm col}$ is initially high during the early stage of the outburst and then gradually decreases. This phenomenon can be attributed to the fact that the disc is still condensing from the thermal corona and the disc density has not yet reached equilibrium. As the disc density increases and approaches equilibrium, $f_{\rm col}$ decreases rapidly. In addition, there are also reports indicating that $f_{\rm col}$ is positively correlated with the non-thermal flux \citep[EXO 1846--031;][]{Ren2022}, and inversely correlated with the disc flux \citep[][Ren et al. 2023 submitted to MNRAS]{Dunn2011}. Our results indicate that the gradual decrease of $f_{\rm col}$ is anti-correlated to the gradual increase in disc flux and positively correlated with the gradual decrease of the non-thermal flux and accretion rate (see Fig.~\ref{fig:flux}).

\subsection{Origin of the Flare}

During the flare, the HR is harder compared to the non-flare epochs. This suggests that the flare is mostly due to the non-thermal component. Furthermore, our spectral analysis reveals that the speed of the flare during this epoch is about 0.5\,c. This implies that the flare is unlikely to originate from the outflow, as outflows typically exhibit much lower speeds, of several hundred or thousand kilometers per second \citep[e.g.,][]{King2012, King2013b}. Moreover, it is important to highlight that the outflow primarily influences the evolution speed of the accretion rate rather than the magnitude of the accretion rate itself \citep[e.g.,][]{Avakyan2023}. 

In addition, although \textit{NICER} did not perform observations during the occurrence of the flare, it detected a quasi-simultaneous weak type-B QPO signal approximately three hours after the X-ray flare, which lasted for about a day. Typically, a type-B QPO is linked to the presence of ejecta, as indicated in prior studies \citep[e.g.,][]{Fender2006, MillerJones2012, Zhang2023, Ma2023}. It is worth noting that \citet{Russell2019} reported the observation of radio ejecta using data from the Australia Telescope Compact Array (ATCA). Their analysis, based on a decelerating motion model, suggests that the possible launch time of the ejecta is MJD $58024.1^{+2.6}_{-3.2}$. In this work, we establish a correlation between the launch time of radio ejecta, the observed onset time of type-B QPO, and the appearance time of the X-ray flare. Similar investigations have been conducted by \citet{Walton2017} using \textit{NuSTAR} data to analyze six strong flares during the 2015 outburst of V404 Cyg. Their findings indicate that these flares are associated with transient jet activity, as evidenced by the flux-resolved properties of the flares and the simultaneous onset of radio activity.

While numerous reports have extensively covered ejecta emission and state transition from the HIMS to the SIMS, specifically referred to as ``type-B appearance'' \citep[e.g.,][]{Fender2009, MillerJones2012, Ma2023}, little attention has been given to the duration of type-B QPOs following the emergence of ejecta. In MAXI J1348--630, an ejecta was observed with MeerKAT and ATCA \citep{Carotenuto2021}. A detailed discussion in \citet{Carotenuto2022} suggested that the jet was launched one day prior to the appearance of type-B QPOs, and that these type-B QPOs persisted for about $1.7$ days \citep{Zhang2020, Zhang2021, Liu2022}. In addition, in the case of MAXI J1820+070, the ejecta launch time, obtained using VLBI data with the dynamic phase center tracking technique \citep{Wood2021}, coincided with the emergence of type-B QPOs \citep{Homan2020, Ma2023}. These type-B QPOs lasted for about $3.5$\,hr until their significance decreased beyond detectable levels \citep{Homan2020, Ma2023}. In our analysis, we demonstrate that the launch time of ejecta is likely correlated with the onset of type-B QPO (the simultaneous X-ray flare was not captured by \textit{NICER}), and that type-B QPO persists for up to one day. 

Finally, it is noteworthy that no additional non-thermal components were detected in the energy spectrum during the flare period. This observation aligns with expectations, given the relatively weak nature of the type-B QPO during this phase, making that the difference in energy spectra may not be statistically distinguishable. In addition, a comprehensive investigation into changes in the energy spectrum concomitant with the appearance and disappearance of type-B QPOs across BHXBs, including GX 339--4, MAXI J1820+070, and MAXI J1348--630, was undertaken by \citet{Yang2023}. Their results revealed similarities in the energy spectral shapes between the presence and absence of type-B QPOs. Employing a preliminary estimation of the rms amplitude of the QPO, they ascertained that there was no need for supplementary non-thermal components within the spectrum. This also corroborates the consistency of the energy spectral model between flare and non-flare periods in MAXI J1535--571.

\section{Conclusions}
\label{sec:con}

We analyze a \textit{NuSTAR} and a quasi-simultaneous \textit{NICER} observation of MAXI J1535--571 during the SIMS, focusing on a 2.5-ks X-ray flare observed by \textit{NuSTAR}. We study the spectral and timing properties of the source during this event, aiming to explain the origin of the flare. Our key findings are: 

\begin{itemize}
\item[1.]
We found a high black-hole spin, $a=0.97_{-0.10}^{+0.02}$, and a low source inclination, $i\sim24^{\circ}$. The small inclination suggests the probable presence of a warped disc in MAXI J1535--571.
\end{itemize}

\begin{itemize}
\item[2.]
As MAXI J1535--571 transitions to the HSS, the apparent disc radius exhibits an unusually increasing trend with increasing disc flux, which we attribute to a decrease of the color correction factor.
\end{itemize}

\begin{itemize}
\item[3.]
Our analysis suggests that the X-ray flare is associated with the discrete jet: The flare comprises a hard component with a high velocity of approximately 0.5 c. Additionally, timing analysis reveals the appearance of a quasi-simultaneous weak type-B QPO and the launch of a radio ejecta coincides with the X-ray flare.
\end{itemize}




\section*{Acknowledgements}

This work is supported by the National Key R\&D Program of China (2021YFA0718500). We acknowledge funding support from the National Natural Science Foundation of China (NSFC) under grant No. 12122306, and U2031205, the CAS Pioneer Hundred Talent Program Y8291130K2, and the Scientific and Technological Innovation project of IHEP Y7515570U1. MM acknowledges the research programme Athena with project number 184.034.002, which is (partly) financed by the Dutch Research Council (NWO).


\section*{Data Availability}

The data used in this article are available in the HEASARC data base (\url{https://heasarc.gsfc.nasa.gov}). The corner plots were created using {\sc tkXspecCorner} (\url{https://github.com/garciafederico/pyXspecCorner}).



\bibliographystyle{mnras}
\bibliography{ref} 




\appendix

\section{Corner figures}

\begin{figure*}
\centering
\includegraphics[width=\textwidth]{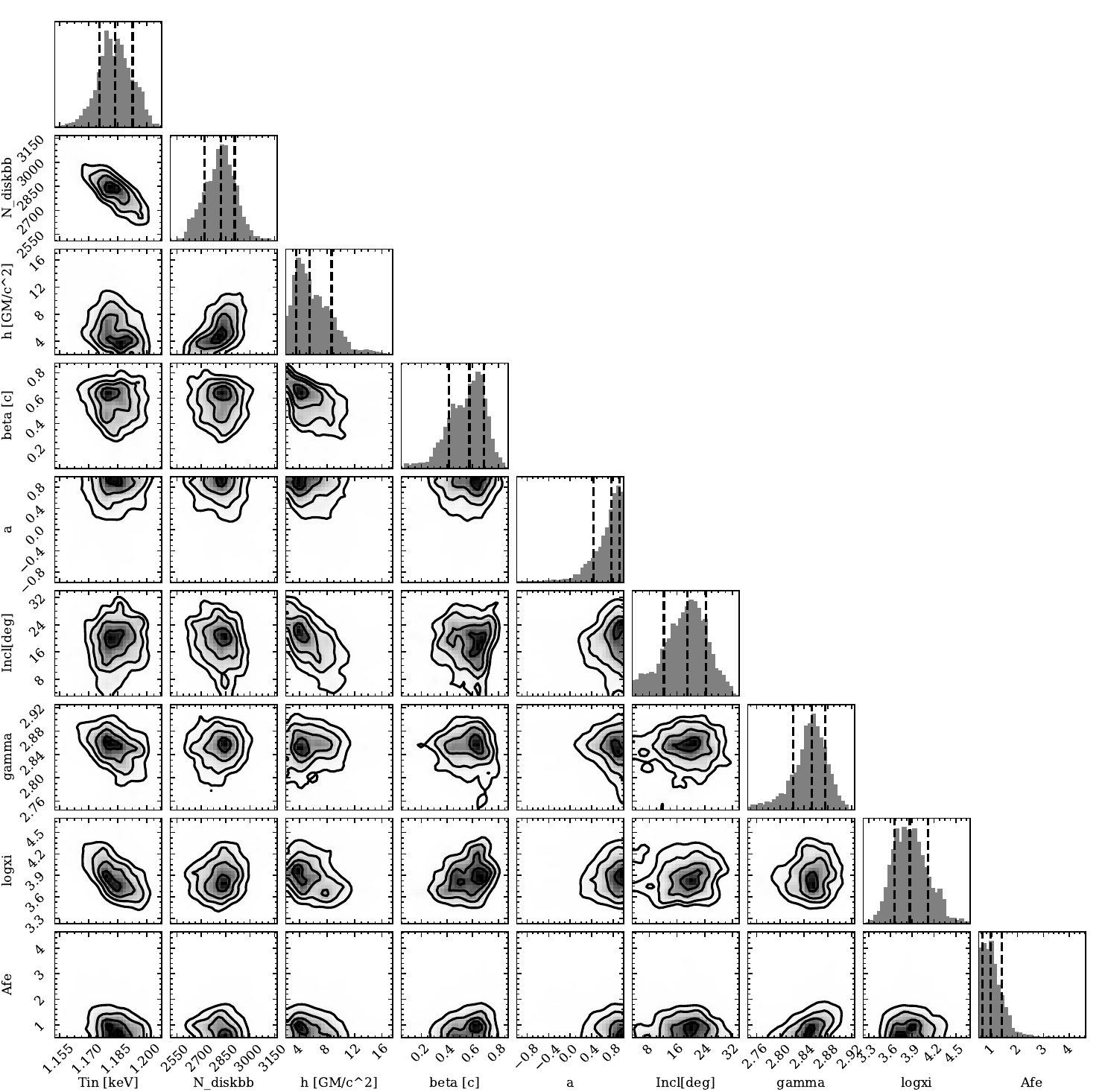} \\ 
\caption{Representative corner plot of the parameters of MAXI J1535--571 for epoch 1 is determined through individual fitting.. The contours indicate the 1, 2 and 3-$\sigma$ levels for two parameters, while the vertical lines indicate the 1-$\sigma$ confidence range for one parameter.
\label{fig:Cont_epoch1}}
\end{figure*}

\begin{figure*}
\centering
\includegraphics[width=1.05\textwidth]{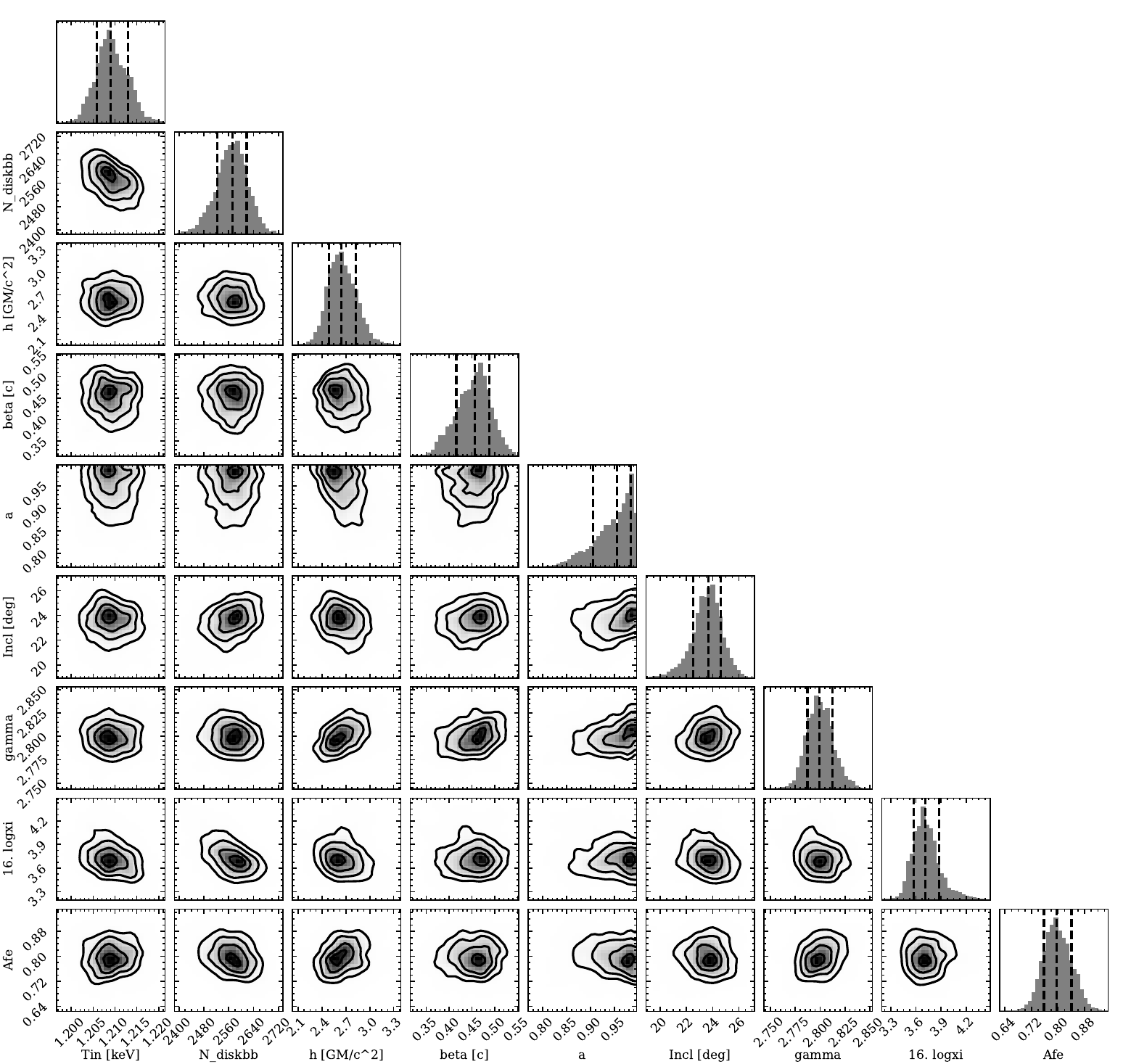} \\ 
\caption{Same as Fig.~\ref{fig:Cont_epoch1} for the joint fitting.}
\label{fig:Cont_joint_epoch1}
\end{figure*}

\bsp	
\label{lastpage}
\end{document}